\documentclass{aa}
\usepackage{amssymb}
\usepackage{amsmath}
\usepackage{xspace}
\usepackage{graphicx}

\def\co{$^{59}$Co }
\def\ni{$^{59}$Ni }
\usepackage{natbib}

\def\gr{$\gamma$-ray}

\begin{document}
\title{Abundances of $^{59}$Co and $^{59}$Ni in the cosmic ray flux}
\author{A.Neronov, G.Meynet}
\institute{
Astronomy Department, University of Geneva, Ch. d'Ecogia 16, 1290, Versoix, Switzerland \\
}

\abstract
{Two main hypotheses for the origin of Galactic cosmic rays are the "supernova" and "superbubble" origin hypotheses. }
{
We analyse the evidence for the superbubble hypothesis provided by the measurements of the relativive abundances of isotopes of cobalt and nickel in the cosmic ray flux. }
{ 
We compare the measured upper limit on the abundance of \ni in the cosmic ray flux with the  \ni abundance predictions of the up-to-date stellar evolution models. Non-detection of \ni in the cosmic ray flux has previously been attributed to a large time delay of the order of $10^5$~yr  between the moment of supernova explosion and the onset of particle acceleration process. This large time delay was considered as an argument in favour of  the "superbubble" scenario. }
{
We show that the recent calculation of the \ni yield of massive stars, which takes into account the initial mass range up to 120 solar masses and includes stellar rotation, results in prediction of low \ni abundance relative to its decay product \co. The predicted abundance is consistent with the upper bound on \ni abundance in the cosmic ray flux for the supernova parameters assumed. This result removes the necessity of decay of \ni in the time interval between the supernova explosion and the onset of acceleration process and restores the consistency of measurements of \ni / \co abundances with the "supernova" hypothesis of the CR origin.  }
{} 


\maketitle

\section{Introduction}

Observations of \gr\ emission from star forming galaxies  demonstrate that acceleration of cosmic rays (CRs) is a by-product of star formation activity \citep{fermi_starforming}. Two alternative possibilities for the star-formation induced CR acceleration could be considered.   CRs injection could occur in supernovae and in objects related to the supernovae \citep{baade34}, via shock acceleration \citep{krymskii77,bell78,blandford78,drury83,bell01} occurring in supernova remnants, pulsar wind nebulae and/or gamma-ray bursts (see \citet{blasi_review,grenier15} for recent reviews). Evidence for such a scenario is found in \gr\ observations which reveal the presence of high-energy protons in supernova remnants \citep{fermi_snr}.  Alternatively, CR acceleration might occur in the superbubbles formed by collective effects of multiple supernovae and winds from massive star \citep{bykov92,higdon98,parizot04,binns05,bykov14}. Evidence for such a scenario could also be found in the \gr\ data, which reveal injection of cosmic rays in the nearby superbubble in Cygnus X region \citep{cocoon}. 

Chemical composition of the CR flux  carries valuable information on the CR sources \citep{wiedenbeck81,maeder93,binns05,prantzos12}. The abundances of different nuclei in the CR flux deviate from the abundances in the Solar system. This is expected in both the individual supernova and the superbubble scenarios. 

In the case of supernovae,  the ejecta and the circumstellar medium around the supernova are enriched with the heavy elements produced at different stages of evolution of the massive star. If the acceleration of CRs occurs sufficiently early after the supernova explosion, at the free expansion stage and/or at the beginning of the Sedov-Taylor phase, the abundance pattern of the accelerated CRs is also expected to be enriched with heavy elements. To the contrary, if the acceleration occurs late in the Sedov-Taylor phase, the medium which provides the source of the accelerated particles has already the composition close to that of the ISM and CRs produced at this stage would have the abundance pattern similar to that of the ISM.  The details of the enrichment are sensitive to the dynamics of the propagation of the forward shock of the expanding supernova shell into the cavity carved by the wind produced by the progenitor massive star \citep{prantzos12}. 

In a similar way, the composition of the medium inside the superbubble changes from the standard ISM composition at the onset of the star formation to the composition enriched with heavy elements injected by the massive star winds and supernovae \citep{casse82,binns05}. Composition of the CRs accelerated at the multiple shocks produced by the supernovae and stellar wind bubbles would also change in the course of evolution of the superbubble. 

Particularly interesting indications about the details of the CR acceleration process might be provided by the ratios of abundances of isotopes such as the "anomalous" $^{22}$Ne/$^{20}$Ne \citep{mewaldt80,casse82,binns05} and $^{59}$Ni/$^{59}$Co ratios \citep{wiedenbeck99,israel05}. Information provided by the isotope ratios is potentially free from uncertainties related to the uncertainty of the mechanism of injection of charged particles into the acceleration site. The injection efficiency most probably depends on the characteristics of particular elements, such as  mass, charge, the first ionisation potential, volatility etc \citep{ellison97}. Dependence of the injection efficiency on these parameters might lead to the distortion of the abundance pattern of elements in the CR flux, as convincingly demonstrated e.g. for the first ionisation potential correlation with the elemental abundance \citep{wiedenbeck07}.

Nickel isotope $^{59}$Ni is unstable and decays into $^{59}$Co on the decay time scale $t_d=7.6\times 10^4$~yr through the electron capture. However, it could not decay in this way if accelerated and converted into a cosmic ray soon after the supernova explosion. Non-detection of $^{59}$Ni in the cosmic ray flux apparently indicates that the $^{59}$Ni had a possibility to decay before the cosmic ray acceleration process started. This fact supports the hypothesis that the CR acceleration process does not occur in young supernova remnants and is instead efficient only in the superbubble environment. 

The conclusion on the decay of $^{59}$Ni before the onset of the cosmic ray acceleration process is based on the comparison of the upper bound on the  $^{59}$Ni/$^{59}$Co ratio in the CR flux to the $^{59}$Ni and $^{59}$Co yields of massive stars ending their life in supernova explosions. The predictions for the abundances of these isotopes are taken from the calculations of \citet{woosley95}. These calculations describe the elemental yields produced throughout the life cycle of stars including the moment of supernova explosion. The predicted ratio $f_{Ni}=^{59}$Ni/($^{59}$Co+$^{59}$Ni) varies in the 0.2-0.9 range, with the median value $\simeq 0.5$. Comparison of this value with the CR measurements which constrain $f_{Ni}<0.18$ in the CR flux has led  \citet{wiedenbeck99} to the conclusion that  $^{59}$Ni decays before the onset of the acceleration process.  

In what follows we notice that the models of \citet{woosley95} did not extend to the entire relevant range of the initial masses of massive stars. \citet{wiedenbeck99} have used the results of calculations of \citet{woosley95}  up to the initial mass $25M_\odot$ to estimate $f_{Ni}$. We notice that stars of higher masses, being less abundant, could still provide a sizeable amount of $^{59}$Co and $^{59}$Ni. We use more recent calculations of the elemental yields by \citet{chieffi13} to show that account of these heavier stars can result in the model prediction of $f_{Ni}$ which is consistent with the upper bound found on $f_{Ni}$  in the CR flux. 

\section{$^{59}$Co and $^{59}$Ni yield from massive stars and supernovae}

Different isotopes of the iron group nuclei, including cobalt and nickel could be produced at different stages of stellar evolution, starting from the helium hydrostatic burning stage up to the explosive nucleosynthesis at the moment of the supernova explosion.
It happens that a significant amount if not most of $^{59}$Co is synthesized during the hydrostatic burning phases through neutron captures by iron-peak elements, the so called s-process, while $^{59}$Ni is mainly made during the very last stages of the evolution of massive stars, during Si-burning. This can be seen for instance by looking at Table~9 in \citet{woosley95}, where the quantities of $^{59}$Co and of $^{59}$Ni in the star are given at different stage of the evolution of a solar metallicity 25 M$_\odot$ model. Typically, the amount of $^{59}$Co at the end of the core He-burning stage is already not far from the final amount ejected, while that of $^{59}$Ni is still very low at that stage and takes its final value mainly as a result of the explosive nucleosynthesis.
Thus, the quantities ejected of these two isotopes depend on very different physical ingredients of the models. The ejected mass of $^{59}$Co is quite sensitive to the mass of the
helium core (larger amounts for larger He-core masses), but depends weakly on the physics of the explosion. In contrast, the amount of $^{59}$Ni does not depend much on the size of the He-core but is very sensitive to the physics of the explosion. Of course many other parameters enter into this game:
the initial mass of the star, its rotation velocity, metallicity, magnetic field strength, nuclear reaction rates etc. Uncertainty of these parameters unavoidably introduces uncertainties to the calculation of the pre-supernova relative abundance of $^{59}$Co and $^{59}$Ni.  The amounts of $^{59}$Co and $^{59}$Ni produced also depend on the parameters for which no direct observational data is available and which could be assessed only through numerical simulations, such as e.g. the time-dependent neutron excess. Finally, calculations of elemental yields of massive stars include phenomenological parameters which parametrize our uncertainty of the knowledge of the physical mechanisms behind certain phenomena. An example of such a parameter is the "mass cut" which sets a boundary between the material in the interior of the star which is ejected by the supernova explosion and the material which ultimately forms the compact remnant of the explosion. Increase of the computing power opens possibilities for gradual  improvement of precision of the stellar evolution models and account of larger number of relevant parameters regulating the evolutionary path stars  \citep[see][for a review]{woosley02}. 

\begin{figure}
\includegraphics[width=\linewidth]{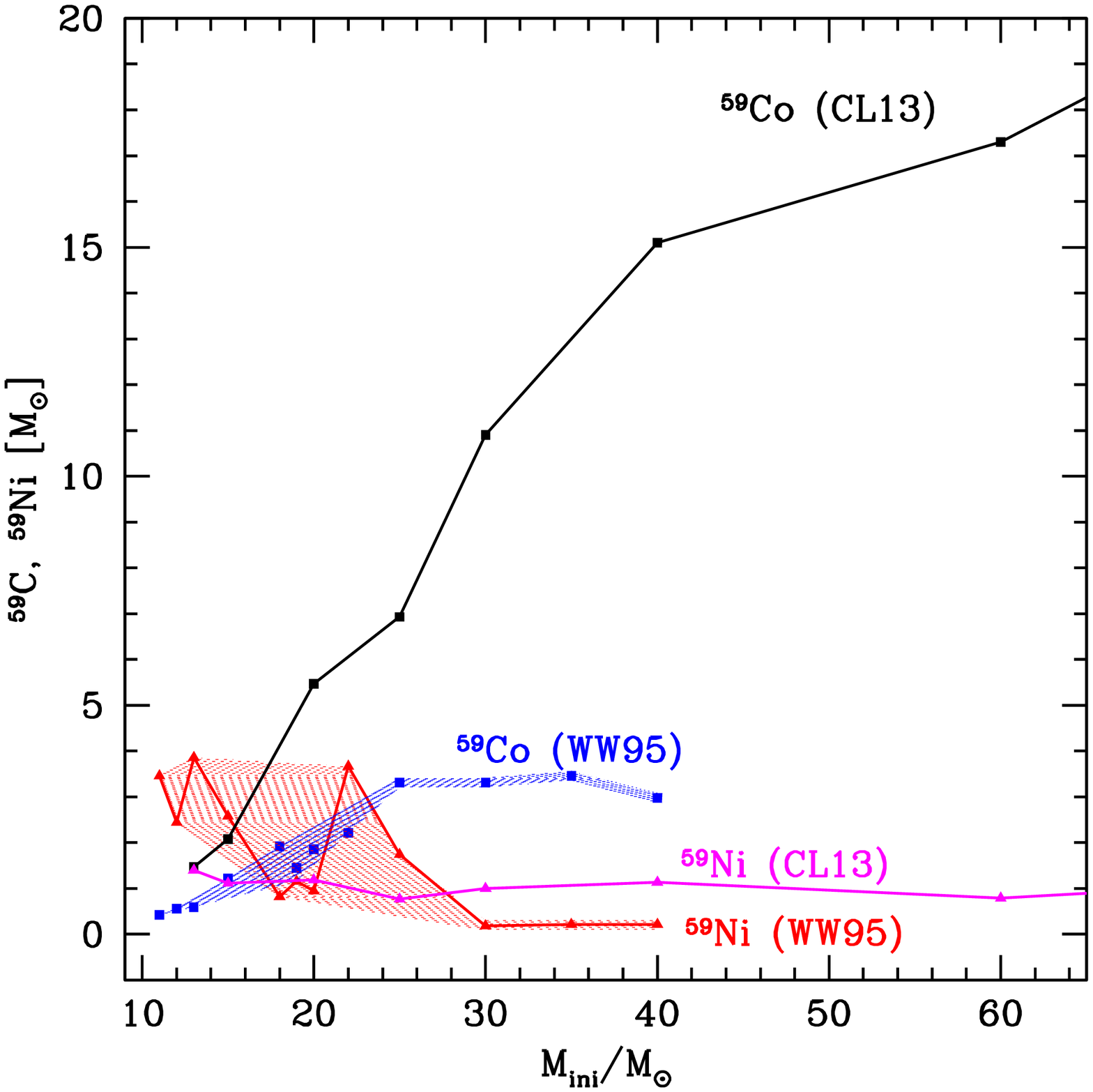}
\caption{
Ejected masses of $^{59}$Co and $^{59}$Ni by star of different initial masses and computed by different authors: WW95 stands for  \citet{woosley95} and CL13 for  \citet{chieffi13}. Both models are for solar metallicity.
The models from \citet{woosley95} correspond to their A series, the models of  \citet{chieffi13} correspond to their rotating models. Note that in both cases, the yields for the radionuclides are given 2.5 10$^4$ sec. after the explosion.
That means that $^{59}$Ni had no time to decay (half-life is about 10$^5$ years). The masses in $^{59}$Ni of \citet{woosley95} present a kind of sawtooth behaviour 
below 25 M$_\odot$ (see the red continuous lines), a hatched area covers that zone to guide a bit the eye. This was done also for the mass of $^{59}$Co from  \citet{woosley95} although here the variations are much smaller.}
\label{fig:nico}
\end{figure}

The most recent calculation of the elemental yields of massive stars which includes the predictions of the \co and \ni isotopes is the set of models of \citet{chieffi13}. These yields are compared to those of  \citet{woosley95} in Fig.~\ref{fig:nico}.
A few interesting points can be noted.
\begin{itemize}
\item The yields in $^{59}$Co of  \citet{chieffi13} are significantly larger than those of \citet{woosley95}, although they present a qualitative similar dependence with the initial mass
(increasing with it). 
These differences are at least in part due to the fact that the models of \citet{chieffi13} have, for a given initial mass, a larger He-core than the models of  \citet{woosley95}.
\item In both series of models  $^{59}$Co is made essentially during the hydrostatic burning stages. In the case of the models by \citet{chieffi13}, this conclusion is based
on a comparison we performed between their $^{59}$Co yields and those obtained recently by \citet{frishknecht16} who just computed the contribution of the
s-process occurring during the hydrostatic burning stages. If we superpose $^{59}$Co yields for the masses between 15 and 40 M$_\odot$ with and without rotation
of these two papers, we obtain a nearly perfect match. This confirms that indeed $^{59}$Co is made mainly by the s-process.
\item For $^{59}$Ni, the yields of \citet{chieffi13} are generally smaller than those of \citet{woosley95} in the mass domain between 13 and 25 M$_\odot$ and 
larger for masses above 25 M$_\odot$. This is quite consistent with the fact that below 25 M$_\odot$, the mass cuts (the kinetic energy of the explosion) in the models of  \citet{chieffi13} are larger  (smaller) than in the models by \citet{woosley95} and the reverse for the masses above 25 M$_\odot$. 
As mentioned above, the $^{59}$Ni yields are very dependent on the physics of the explosion. 
\item In the mass domain between 11 and 25 M$_\odot$, the models by \citet{woosley95} indicate strong variations of the $^{59}$Ni yields with the initial mass. Typically, the yields for the
20, 22 and 25 M$_\odot$ are respectively (in units of 10$^{-4}$ M$_\odot$) 0.95, 3.67 and 1.74. It is difficult to give a very clear physical reason for that behavior. We can just note that
the mass of $^{56}$Ni shows also strong variations (0.088, 0.205 and 0.129 M$_\odot$). On the other hand, the models by   \citet{chieffi13} give for the 20 and 25 M$_\odot$ rotating models
(they have no predictions for a 22 M$_\odot$ model) yields of $^{59}$Ni equal to 1.19 and 0.77 10$^{-4}$ M$_\odot$ and yields in $^{56}$Ni equal to 0.1 M$_\odot$ (actually in these models
the mass cut has been chosen to obtain such an amount of $^{56}$Ni). 
\end{itemize}

\begin{figure}
\includegraphics[width=\linewidth]{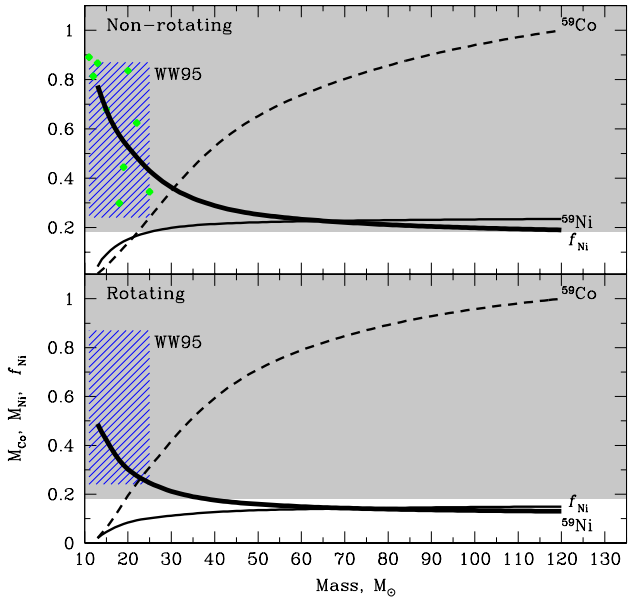}
\caption{Cumulative mass yields $M_{^{59}Co}$ (thin dashed lines) and $M_{^{59}Ni}$ (thin solid lines) as a function of the maximal initial mass for the models resulting in $0.1M_\odot$ of $^{56}$Ni. The masses are normalized on the total \co mass. Thick solid line shows $f_{Ni}$, also as a function of the maximal initial mass.   Green points in the upper panel indicate $f_{Ni}$ obtained from model calculations of \citet{woosley95} in the mass domain between 11 and 25 M$_\odot$, that is in the mass domain that was considered in the work by \citet{wiedenbeck99}.  }
\label{fig:CoNi}
\end{figure}

We use the models by \citet{chieffi13}  for our calculation of $f_{Ni}$.  
The top panel of Fig. \ref{fig:CoNi} shows the cumulative  masses of $^{59}$Co and $^{59}$Ni produced by a population of massive stars with Salpeter initial mass function \citep{salpeter55} in the mass range between $13M_\odot$   \citep[the smallest mass of the models of][]{chieffi13} up to the maximal mass shown as the $x$ axis value.   The values correspond to the time moment $2.5\times 10^4$~s after the explosion, i.e. much before the decay of \ni into \co. One could see that the $^{59}$Co mass accumulates linearly at a steady rate up to the stellar mass range $\simeq 40M_\odot$ and at somewhat lower rate in the mass range $M>40M_\odot$. Still, consideration of the full dynamic range of initial stellar masses is important for a fair judgement of the amount of $^{59}$Co produced. 

Figure \ref{fig:CoNi} shows that a large part of $^{59}$Ni is produced by the stars in the mass range up to $(20-30)M_\odot$. This explains the decrease of the Nickel-59 relative abundance $f_{Ni}$ with the increase of the maximal initial stellar mass. Integration of the  \co and \ni yields over the entire mass range results in the 
\begin{equation}
f_{Ni}=0.19 \mbox{\ \ \  (non-rotating)}
\end{equation}
\ni fraction.

An essential difference of the models of \citet{chieffi13} from the models of \citet{woosley95} is the account of the rotation in the calculation of the stellar evolution. The bottom panel of Fig. \ref{fig:CoNi} shows the \co and \ni yields of stars rotating with initial equatorial velocity 300~km/s. In spite of similar qualitative behaviour of the cumulative yields of \co and \ni, the rotating models predict systematically lower \ni to \co fractions, which gives 
\begin{equation}
f_{Ni}=0.13 \mbox{\ \ \  (rotating)}
\end{equation}
after integration over the entire initial stellar mass range.

Similar integrations performed with the yields of \citet{woosley95} provide values of $f_{Ni}$ between 0.6 and 0.7 depending whether
the integration is made between 11 and 40 M$_\odot$ or between 11 and 25 M$_\odot$.

Lower values of  $f_{Ni}$ from the models of  \citet{chieffi13}  result mainly because the high mass range show very low $^{59}$Ni to $^{59}$Co fraction, while this mass domain
still contributes significantly to $^{59}$Co. As we shall see this has important consequences for interpreting the observed upper limit of $f_{Ni}$ in cosmic rays (see next section).

\begin{figure}
\includegraphics[width=\linewidth]{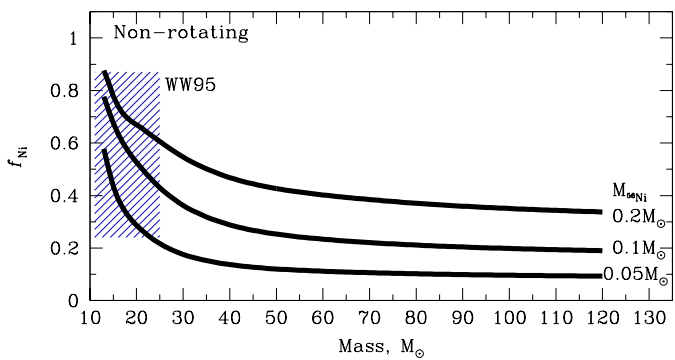}
\caption{ $^{59}$Ni fractions as a function of the assumed $^{56}$Ni yield for non-rotating star models.   }
\label{fig:CoNi_masscuts}
\end{figure}

An additional uncertainty of $f_{Ni}$  estimate is introduced by the uncertainty of the position of the "mass cut" separating the ejecta from the collapsar. The predicted amounts of both $^{56}$Ni and $^{59}$Ni increases if the mass cut is moved deeper in the iron elements rich core. As a result, $f_{Ni}$ increases with the increase of the assumed $^{56}$Ni yield. This is illustrated by Fig. \ref{fig:CoNi_masscuts} which compares the dependence of $f_{Ni}$ on the maximal initial mass calculated for different $^{56}$Ni yields. One sees that a variation of $^{56}$Ni yield from $0.05M_\odot$ to $0.2M_\odot$ \citep[around the reference value $0.1M_\odot$ assumed by][]{chieffi13} results in changes of $f_{Ni}$ in the 0.1-0.3 range. Observationally, the yield of $^{56}$Ni is observed to be about $0.1M_\odot$ for stars with progenitor masses $\lesssim 25M_\odot$ and is perhaps spread over more than an order of magnitude for higher mass progenitors \citep{nomoto}.

What are the effects that could modify $f_{Ni}$ in addition of those produced by considering different sets of stellar models?

A main effect comes from the uncertainties about the fate of stars producing a black hole at the end of their evolution. Do these stars swallow all their mass into the newly formed black hole 
or do they still expel some material in a supernova event? In case the transition between neutron star and black hole
production would be around 25 M$_\odot$ \citep{Heger03}, and that, when a black hole is formed, no matter ejection occurs, then 
the integration for obtaining $f_{Ni}$ should be made only in the mass range below 25 M$_\odot$ as was done in  \citet{wiedenbeck99}.
We see that, in that case, even using the models by  \citet{chieffi13}, would produce high  $f_{Ni}$ values. This would then
support the view that some time delays took place between the ejection of $^{59}$Ni ejection and its acceleration into the cosmic rays.
Although such a possibility cannot be totally discarded at the moment, there are a few arguments indicating that reality might be more complexe:
\begin{itemize}
\item First, the domain of initial masses giving neutron stars and black holes is not so well defined. Hydrodynamical simulations in 1 D by \citet{Ugliano12, Ertl15}
obtain successful SN explosions with neutron star formation for some models with initial masses between
20 and 40 M$_\odot$, and even up to 120 M$_\odot$, while failed explosion with BH formation seem also possible for progenitors below 20 M$_\odot$!
\item Second, there are observations of the surface abundances of stars near a black-hole, showing clear evidences for the star having been enriched 
by the ejecta of the SN that occurred at the time of the black hole formation. This indicates that,
at least in some cases, a BH can form with a SN explosion \citep{Ga99, Gon04, Gon08, Gon11}.
\item At solar metallicity, the most massive stars lose anyway mass by stellar winds. Only $^{59}$Co has a chance of being ejected by the stellar winds (when the star
is a Wolf-Rayet star of the type WC, {\it i.e.} where the He-burning products appear at the surface). Thus, in case these stars
would eject no material at the time of the supernova event, they would only contribute to the enrichment in $^{59}$Co. 
It remains however to be checked whether the amounts of $^{59}$Co ejected in that way are sufficient to lower $f_{Ni}$.
 \end{itemize}
The above arguments cannot be taken as proofs that we must make the integration over the whole mass range in the way we did above, but at least it shows that restricting its computation to only the mass domain below 25 M$_\odot$ is probably not realistic.

\section{Implications for the CR origin problem}

Recent calculations of the yields of \co and \ni based on the detailed modelling of the pre-supernova evolution of massive stars over a wide energy range from $13M_\odot$ to $120M_\odot$ with account of rotation \citep{chieffi13} differ from previous calculations of \citet{woosley95}. In particular they result in a systematically lower \ni  to \co fraction after the moment of supernova explosion (for the reference models normalised on $0.1M_\odot$ $^{56}$Ni yield). 

This fact has important consequences for the physics of Galactic CRs. 
The absence of \ni in the CR flux has been considered as an evidence for a large time delay between the moment of supernova explosion and the onset of particle acceleration. This large time delay of the order of $\sim 10^5$~yr was necessary to allow \ni to decay to \co via electron capture reaction. After such time delay, supernovae exploding in the massive star formation regions could not be considered as isolated objects. On such time scale interactions of supernova shell with the superbubble environment have to be taken into account. Therefore, large time delay between the explosion and the moment of particle acceleration has been considered as an argument in factor of the superbubble hypothesis of the CR production. 

The revised calculation of the \ni and \co yield removes requirement of \ni decay prior to the acceleration. The non-detection of \ni in the CR flux might be explained by the relatively small amount of \ni ejected by the ensemble of supernovae produced by stars with different initial masses and rotation velocities. This removes the contradiction between the supernova hypothesis of CR production and the measurements of \ni to \co ratio.

Note that estimating $f_{Ni}$ as done above by integrating over the massive star range, implicitly assumes that the material of the cosmic rays is made from a well mixed reservoir of many different core collapse supernova events. This is quite in line with the fact that the composition of the cosmic rays is indeed quite similar, with a few exceptions (see Sect.~1), to the interstellar medium abundances whose abundance also results, at least in part, from the contributions of many core collapse supernovae.

The above discussion tells us that present day observations are compatible with the acceleration of fresh ejecta of core collapse supernova material into the cosmic rays, provided  the recent yields from the rotating stellar models of \citet{chieffi13} are used, that the whole mass domain from 13 up to at least 45-50 M$_\odot$  contribute and that for computing the SN contribution, mass cuts giving $0.1M_\odot$ of $^{56}$Ni or lower are used. 

Now what does happen when the yields of type Ia supernovae are accounted for? These supernovae may contribute up to about  two thirds of the iron peak element content in the present day Universe \citep{Matt86}.  According to the work by \citet{Trava04}, type Ia's produce typically as much $^{59}$Co as $^{59}$Ni in amounts equal to 5-8 10$^{-4}$ M$_\odot$. In this case, the iron peak elements accelerated into the cosmic rays would  also be made of two thirds of type Ia material  this would give $f_{Ni}$ a bit smaller than 0.5 using the yields by  \citet{chieffi13} and a bit larger than 0.5 using the yields by \citet{woosley95}. This would again support the view that some time delay has occurred between the ejection of $^{59}$Ni and its acceleration  into the cosmic rays. 

However, we must keep in mind several differences between the core collapse supernovae and the type Ia supernovae. 

First, Type Ia  and core collapse supernovae are distributed differently in the Galaxy. The  core collapse supernovae follow the star formation rate. Their density is enhanced in the regions of active star formation, such as the Galactic arms.  Passage of Galactic arm  induces an enhancement of the core collapse supernovae contribution to the cosmic ray population on the time scale of cosmic ray residence in the Galactic disk,  $\sim 10^7$~yr \citep{shaviv03,overholt09,piran13}. To the contrary, Type Ia supernovae rate does not follow temporal and spatial variations of the star formation rate. Instead, it is sensitive to the time integral of the star formation rate averaged over the Galactic volume. Even if the injection efficiency of cosmic rays by the core collapse and Type Ia supernovae is the same, relative importance of the Type Ia and core collapse supernovae contributions to the cosmic ray flux could vary from place to place depending  on the details of the recent massive star formation rate around the point of measurement \citep{neronov12,kachelriess15}.  

Next, type Ia supernovae are believed to leave no remnant, while core-collapse supernovae do.  In case supernova remnants play an important role in the acceleration mechanism \citep{neronov_semikoz},  the core collapse supernovae might be more efficient accelerators due to the presence of the remnants. 

Finally, the core collapse supernova initially expands in the bubble created by the wind of the progenitor star, while the Type Ia supernova shell could expand in the interstellar medium already at the early stages of its evolution. In this case the forward shock of the Type Ia supernova sweeps the material with the composition of the interstellar medium. If acceleration at the forward shock provides a dominant contribution to the cosmic ray yield of the Type Ia supernovae \citep{warren} composition of the cosmic rays originating from this shock is not directly related to the elemental yield of the supernova. The interstellar medium feeding the shock contains no $^{59}$Ni so that even cosmic rays from the Type Ia supernovae could be characterised by low $f_{Ni}$. 

Actually, the last argument could also be valid for the core collapse supernovae.  Assumption that cosmic ray production in these supernovae is also dominated by acceleration at the forward shock would just invalidate the "naive" assumption that cosmic ray elemental yield of a supernova just repeats the elemental yield of the star after the explosion implicitly adopted by \citet{wiedenbeck99} and also in the previous sections of this paper. Instead, the composition of material accelerated at the forward shock at a particular  moment of time (say, at the onset of Sedov-Taylor phase) would be influenced by the composition of portion of the stellar wind swept up by the supernova shell by that time  \citep{prantzos12}.




It is interesting to note that the  \ni fraction predicted by the stellar evolution models of \citet{chieffi13} is close to the measured upper bound on $f_{Ni}$ in the CR flux. This indicates that moderate increase of sensitivity of CR detectors capable to distinguish iron group isotopes, like CRIS \citep{wiedenbeck99} should lead to the detection of the \ni component of the CR flux, if the accelerated particles originate from the core collapse supernova ejecta. Detection of \ni would provide a strong constraint on the mechanism of cosmic ray acceleration, because it would indicate that the supernova ejecta participate in the cosmic ray acceleration process and  limit the time delay between the supernova explosion and the acceleration period to less then $\sim 10^4-10^5$~yr, which is a typical time scale on which the supernova remnants reach the Sedov-Taylor phase \citep{wojter72,vink12}. 


\end{document}